\documentclass[aps,preprint,superscriptaddress,showpacs]{revtex4}
\usepackage{amsmath,bm}
\usepackage{graphicx}

\begin{document}

\title{Monovacancy-induced magnetism in graphene bilayers}
\author{Sangkook Choi}
\affiliation{Department of Physics and Astronomy, Seoul National University, Seoul 151-747, Korea}
\affiliation{Department of Physics, University of California at Berkeley, Berkeley, CA 94720, USA}
\author{Byoung Wook Jeong}
\affiliation{Department of Physics and Astronomy, Seoul National University, Seoul 151-747, Korea}
\author{Seungchul Kim}
\affiliation{Department of Physics and Astronomy, Seoul National University, Seoul 151-747, Korea}
\author{Gunn Kim}
\email[corresponding author. Electronic mail:\ ]{kimgunn@skku.ac.kr} 
\affiliation{BK21 Physics Research Division and Department of Physics, Sungkyunkwan University, Suwon, 440-746, Korea}
\date{\today }

\begin{abstract}

Vacancy-induced magnetism in graphene bilayers is investigated using spin-polarized density functional theory calculations. 
One of two graphene layers has a monovacancy. Two atomic configurations for bilayers are considered with respect to the position of the monovacancy. 
We find that spin magnetic moments localized at the vacancy site decrease by ~10\% for our two configurations, 
compared with the graphene monolayer with a monovacancy. 
The reduction of the spin magnetic moment in the graphene bilayers is attributed to the interlayer charge transfer 
from the adjacent layer to the layer with the monovacancy, compensating for spin magnetic moments originating from quasilocalized defect states.

\end{abstract}

\maketitle
\section{introduction}

Vacancy defects in crystalline solids have been of fundamental interest in materials science and condensed matter physics. 
Of particular interest, both technologically and academically, are the vacancy defects 
in nanostructures of sp$^2$-bonded carbon based nanostructures [1,2]. They exhibit room-temperature (RT) ferromagnetism [3], 
and the underlying physics of their ferromagnetism is different from that of conventional ferromagnetic metals such as iron and cobalt [4,5]. 
Successive progress in the understanding of ferromagnetism in carbon-based nanostructures has been achieved 
by theoretical approaches based on spin-polarized density functional theory [6-9] 
and experiments on irradiated carbon nanostructures [10-12]. 
To illustrate, for a graphene monolayer (a two-dimensional honeycomb lattice of C atoms) with a monovacancy, 
it has been elucidated that spin magnetic moment is localized at the vacancy site and 
that RT-ferromagnetism in the graphene monolayer originates from localized sp$^2$ dangling bond state as well as quasilocalized defect state [4,13]. 
Previous theoretical studies of magnetism in two-dimensional graphitic systems have focused on the graphene monolayer, 
neglecting the interlayer coupling. 
In this paper, we investigate, using first-principles calculations, graphene bilayers with a monovacancy 
to understand the influence of the adjacent graphene layer to the spin magnetic moment of the graphene layer with a monovacancy. 
This result would provide a more realistic understanding of vacancy-induced magnetism in graphene multilayer systems 
including highly oriented pyrolytic graphite (HOPG) [14,15].

\section{computational details}

We perform ab initio calculations based on the density functional theory with spin polarization. 
The wave functions are expanded in the double-$\zeta$ basis set implemented in the SIESTA code [16]. 
Norm-conserving Troullier-Martins pseudopotentials are employed [17]. 
For the exchange-correlation term, we employ the Cerperley-Alder type [18] 
local density approximation with spin polarization (LSDA). 3 $\times$ 3 $\times$ 1 Monkhorst-Pack grids with respect to 1 $\times$ 1 graphene cell 
are used to sample the Brillouin zone and an energy cutoff for real space mesh points is 200 Ry. 
All coordinates are fully relaxed until the forces of each atom are smaller than 40 meV/\AA. 
For the Bernal-stacked (AB-type) [19] graphene bilayer (composed of 255 carbon atoms) with a monovacancy, 
two configurations labeled B$_1$ and B$_2$ are considered in the supercell of 19.68 $\times$ 17.04 (\AA$^2$) with 22 \AA~ 
thickness of vacuum as shown in figure 1. For B$_1$, a monovacancy in the upper layer faces a C atom in the lower layer along z direction. 
For B$_2$, on the other hand, a monovacancy in the upper layer does not face any C atom in the lower layer. 
The C-C bond length in one graphene sheet is ~ 1.42 \AA, and the interlayer distance is $\sim$3.35 \AA. 
With the intention of comparison, a graphene monolayer with a monovacancy and 127 C atoms (labeled M) is also calculated. 
To calculate spin and charge distributions of all carbon atoms in the systmes, we carry out the Mulliken population analysis [20].

\section{results and discussion}

Each upper layer in the relaxed bilayer systems (B$_1$ and B$_2$) has nearly the same atomic arrangement as M in figure 1. 
The sheets are almost planar and the atoms near a monovacancy are displaced owing to the Jahn-Teller effect; 
for B$_1$ and B$_2$, two of the three C atoms around the vacancy are rebonded with bond lengths of 1.78 and 1.79 \AA, respectively. 
The other C atom with a dangling bond protrudes slightly toward the lower layer. 
The unsaturated dangling bond in B$_2$ is somewhat closer to C atoms in the lower layer than that in B$_1$, 
so that the perturbation from the interlayer interaction may result in the total energy of 
B$_2$ lower than that of B$_1$ by 0.03 eV, owing to the weakening of the dangling bond [5]. 
Spin (magnetic moment) densities in both of B$_1$ and B$_2$ configurations show noticeable differences from that in M. 
Figure 2 shows perspective view images of isovalue density surface plots of the spin density distributions ($\rho_{\uparrow}({\bf r})-\rho_{\downarrow}({\bf r}))$ 
in B$_1$, B$_2$, and M. 
Red and blue surfaces correspond to spin densities of + and $-$ 0.01 e /\AA$^3$, respectively. 
For all three configurations, figure 2 shows that spin densities are spatially localized 
at the vicinity of the vacancy site and their distributions have the mirror symmetry with respect to y-axis. 
However, spin density satellites to the vacancy sites in B$_1$ and B$_2$ become smaller than those in M. 
Magnitudes of the magnetic moment in B$_1$, B$_2$, and M manifest this difference more quantitatively. 
Magnetic moments in the upper and lower layers of B$_1$, obtained from the Mulliken population analysis [20], 
are 1.31, 0.04 $\mu _B$, and those of B$_2$ are 1.35 and 0.04 $\mu _B$, respectively. 
In comparison, the magnetic moment in M is 1.52 $\mu _B$. 
It means that the magnetic moments in the upper layer of B$_1$ and B$_2$ decrease by 14 and 11 \%, respectively, 
compared to those in M.
	
To understand the origin of the reduction of the magnetic moment, the interlayer charge transfer was checked. 
We calculated increases in the up-spin, the down-spin, and the total charge in the upper layers 
and their decreases in the lower layers of bilayer systems as listed in table 1; 
the charge increase in the upper layers is calculated by subtracting the total charge in M 
from that in the upper layers, 
and the charge decrease in the lower layers is calculated by subtracting 
the charge in the lower layers from the that of the ideal graphene monolayer with 128 C atoms. 
For the two configurations B$_1$ and B$_2$, 0.18 e and 0.16 e of the total charges are transferred from the lower layer to the upper layer, 
respectively, and most of them occupy only energy levels of down-spin density in the upper layer as shown in table 1. 
These values of the interlayer charge transfer are associated with the decrease of the magnetic moment 
in the upper layer of B$_1$ and B$_2$, 0.21 and 0.17 $\mu _B$, so that we can conclude that the reduction of the magnetic moment 
originates mainly from the interlayer charge transfer from the lower layer to the upper layer 
and their occupation of energy levels of down-spin electrons in the upper layer. 
For the perturbation, the missing C atom at the vacancy site is much more important than the C atom with a dangling bond. 
2p$_z$  orbitals in upper and lower layers overlap and are associated with the interlayer interaction (coupling). 
Because of the missing C atom at the vacancy, the local change in the interlayer interaction occurs. 
For B$_1$ configuration, the vacancy site faces a C atom in the other graphene layer. 
On the other hand, for B$_2$ configuration, the vacancy site faces the center of a hexagon in the other graphene layer. 
Therefore, the charge transfer and the magnetic moment difference in B$_1$ are somewhat larger than those in B$_2$.

Next, we study the influence of the interlayer charge transfer on two origins of monovacancy-induced magnetic moments: 
(1) the localized sp$^2$ dangling bond state from broken $\sigma$ bonds and (2) the quasilocalized defect state from broken $\pi$ bonds. 
To resolve them, we analyze contributions of the orbitals (s, p$_x$, p$_y$, and p$_z$) in the C atoms in B$_1$, B$_2$ and M to the magnetic moments. 
In figure 3(a), it is shown clearly that the difference between the magnetic moments of B$_1$, B$_2$ and M comes from the contribution 
of the p$_z$ orbital. By comparing bilayer systems (B$_1$ and B$_2$) to M, 
we recognize that the reduction of magnetic moments mostly originates from the quasilocalized defect state. 
Figure 3(b) depicts top view images of the magnetic moment density difference, cross-sectioned at the graphene layers with a monovacancy. 
Black-colored grids represent the upper graphene layers. 
The resultant distribution of spin density differences in either case is not localized at the vacancy site but has 
a regular triangular pattern, which corresponds to the typical pattern of magnetic moments 
induced by the quasilocalized defect state [4,21]. 
Therefore, the interlayer charge transfer compensates for the magnetic moment induced by the quasilocalized defect state.

The situation could be more clarified with the density of states (DOS) and the band structure as shown in figures. 4 (a)-(c). 
Figures 4 (a) and (b) show the comparison of band structures between B$_1$ and M and between B$_2$ and M, respectively. 
Left and right columns represent the bands of up-spin electrons and down-spin electrons, respectively. 
The DOS of the graphene bilayer shown in figure 4(c) is calculated from the contribution of its upper layer only. 
Red and black colors represent the contribution of up-spin and down-spin electrons in the bilayers B$_1$ and B$_2$, 
and blue and green represent those in M, respectively. 
For these three systems, we find several similar features. 
The band structures show that localized sp$^2$ dangling bond states are $\sim$0.5 eV below the Fermi energy ($E_F$) 
and quasilocalized defect states are pinned at $E_F$ [4]. 
The DOSs reveal that the magnetic moments stem mainly from the localized sp$^2$ dangling bond state, 
and the contribution of quasilocalized defect state are minor in all three systems. 
However, some differences between bilayer systems and M are also shown in these DOSs and the band structures. 
In contrast to M, the interlayer coupling enhances the localization character of quasilocalized defect states 
near Y point below $E_F$, so that more down-spin states are created below $E_F$ (marked by purple dotted ellipse), 
resulting in occupation of transferred electrons from the adjacent layer mostly in the down-spin quasilocalized defect state. 
The DOS of down-spin electrons shows this tendency more clearly. 
The main peak position in the DOS of the down-spin quasilocalized defect state 
in the bilayer systems moved from above $E_F$ to below $E_F$ with respect to M, and the quasilocalized defect state below $E_F$ 
of the bilayer systems has less dispersive band than that of M, demonstrating an enhancement of the localization character of the bilayer system. 
The band dispersion in the band structure can show localization characters. 
It means that a flat band corresponds to a localized state. 
Therefore less dispersive (flatter) bands are associated with quasi-localized states with heavy effective masses. 
Similar to the case of down-spin electrons, the band structure of the up-spin quasilocalized defect state 
in the bilayer becomes more localized near Y point, compared to M. However, the band of the up-spin quasilocalized defect state near Y is far below $E_F$, 
so that there is little increase in the DOS of up-spin electrons of quasilocalized defect state 
from the interlayer charge transfer. Consequently, enhancement of the localization character of quasilocalized defect states 
near Y point below $E_F$ and occupation of transferred electron from the adjacent layer 
in the down-spin quasilocalized defect state are the major factors that contribute to the reduction of magnetic moments. 
Other bands in the bilayer system are also affected by the interlayer coupling in the graphene bilayer, 
but their contribution to the magnetic moment is negligible.

\section{conclusion}

In conclusion, we demonstrated, by {\it ab initio} calculations, that the interlayer charge transfer 
to down-spin quasilocalized defect state with the enhanced localization character results 
in the reduction of the magnetic moment of the total magnetic moment in graphene bilayers. 
Our study sheds light on the physical behaviors of sp$^2$-bonded carbon structures with vacancies and may lead to a new avenue of carbon-based spintronics.

\section*{acknowledgements}

We acknowledge financial support by the second BK21 project (G.Kim) and Samsung Scholarship from the Samsung Foundation of Culture (S. Choi). 
Computations were performed through the support of the KISTI.

\newpage
\begin{center}
\Large{References}
\end{center}

1. Hashimoto A, Suenaga K, Gloter A, Urita K and Iijima S 2004 Nature (London) 430 870-3

2. Charlier J-C 2002 Acc. Chem. Res. 35 1063-9

3. Coey J M D, Venkatesan M, Fitzgerald C B, Douvalis A P and Sanders I S 2002 Nature (London) 420 156-9
 
4. Kim Y-H, Choi J, Chang K J and Tománek D 2003 Phys. Rev. B 68 125420

5. 2005 Carbon-Based Magnetism: An Overview of Metal Free Carbon-Based Compounds and Materials (Amsterdam: Elsevier) ed Makarova T and Palacio F

6. Yazyev O V and Helm L 2007 Phys. Rev. B 75 125408

7.  Lehtinen P O,Foster A S, Ma Y, Krasheninnikov A V and Nieminen R M 2004 Phys. Rev. Lett. 93 187202

8. Lehtinen P O, Foster A S, Ayuela A, Krasheninnikov A, Nordlund K and Nieminen R M 2003 Phys. Rev. Lett. 91 017202
 
9. Park N, Yoon M, Berber S, Ihm J, Osawa E and Tománek D 2003 Phys. Rev. Lett. 91 237204

10. Banhart F 1999 Rep. Prog. Phys. 62 1181-221

11. Hahn J R and Kang H 1999 Phys. Rev. B 60 6007-17

12. Ohldag H, Tyliszczak T, Höhne R, Spemann D, Esquinazi D, Ungureanu M and Butz T 2007 Phys. Rev. Lett. 98 187204 

13. Pereira V M, Guinea F, Lopes dos Santos J M B, Peres N M R and Castro Neto A H 2006 Phys. Rev. Lett. 96 036801

14. Esquinazi P,  Setzer S, Höhne R, Semmelhack C, Kopelevich  Y, Spemann D, Butz T, Kohlstrunk B and L\"osche M 2002 Phys. Rev. B 66 024429

15. Esquinazi P, Spemann D, Höhne R, Setzer A, Han K-H and Butz T 2003 Phys. Rev. Lett. 91 227201

16. Soler J M, Artacho E, Gale J D, García A, Junquera1 J, Ordej\'on P and Sánchez-Portal D 2002 J. Phys. Condens. Matter 14 2745-79

17. Troullier N and Martins J L 1991 Phys. Rev. B 43 1993-2006

18. Ceperley D M and Alder B J 1980 Phys. Rev. Lett. 45 566-9

19. Bernal J D 1924 Proc. R. Soc. London, Ser. A 106 749-73

20. Mulliken R S 1955 J. Chem. Phys. 23 1833-41

21. Wang Z F, Li Q, Su H, Wang X, Shi Q W, Chen J, Yang J and Hou J G 2007 Phys. Rev. B 75 085424

\newpage

\begin{table}[hd]
\caption{\label{table} Up-spin, down-spin and total charge increases in upper layers and their decreases in the lower layers of B$_1$ and B$_2$.
}

\begin{tabular}{c|c|c|c|c|c|c} 
\hline
Systems & \multicolumn{3}{c}{Charge increase in the upper layer (e)} & \multicolumn{3}{c}{Charge decrease in the lower layer (e)} \\
        & total & up-spin & down-spin &  total & up-spin & down-spin \\
\hline
B$_1$   &  0.18 &  -0.01 &   0.19 &   0.18  &    0.07 &   0.11  \\
\hline
B$_2$   &  0.16  &  0.00 &   0.16 &   0.16  &   0.06 &   0.10  \\
\hline
\end{tabular}
\end{table}

\newpage
\begin{center}
\LARGE{[Figure Captions]}
\end{center}

Figure 1. Relaxed atomic configurations of two kinds of Bernal-stacked graphene bilayers with a monovancy labeled B$_1$ and B$_2$, 
and of graphene monolayers with a monovacancy labeled M. 
Black and gray atoms represent C atoms in upper and lower graphene layers, respectively. Insets show the site at which a C atom 
would be removed to generate a monovacancy before the relaxation.

Figure 2. Perspective view images of isovalue density surface plots of the spin [magnetic moment] density distributions for B$_1$, B$_2$, and M. 
Red and blue colors correspond to the spin [magnetic moment] densities of + and $-$ 0.01 e/\AA$^3$, respectively. 

Figure 3. (a) Contributions of the orbitals (s, p$_x$, p$_y$, and p$_z$) in the C atoms in B$_1$, B$_2$ and M to the magnetic moments 
(b) Top view images of the difference in spin densities in the graphene layers with a monovacancy. 
Black colored grids represent atom positions in the upper layers of B$_1$ and B$_2$.

Figure 4. Band structures of B$_1$, B$_2$, and M in the energy range from $E_F$$-$1 eV to $E_F$$+$1eV in (a) and (b). 
Inset shows the Brillouin zone of the two-dimensional rectangular-shaped superlattice. 
(c) Density of state of upper layers of B$_1$ and B$_2$, and M. 
Purple-dotted regions represent the contribution of the interlayer charge transfer 
to the reduction of magnetic moment in the upper layers of B$_1$ and B$_2$ with respect to M.

\newpage

\newpage
\begin{figure}[t]
  \centering
  \includegraphics[width=11.0cm]{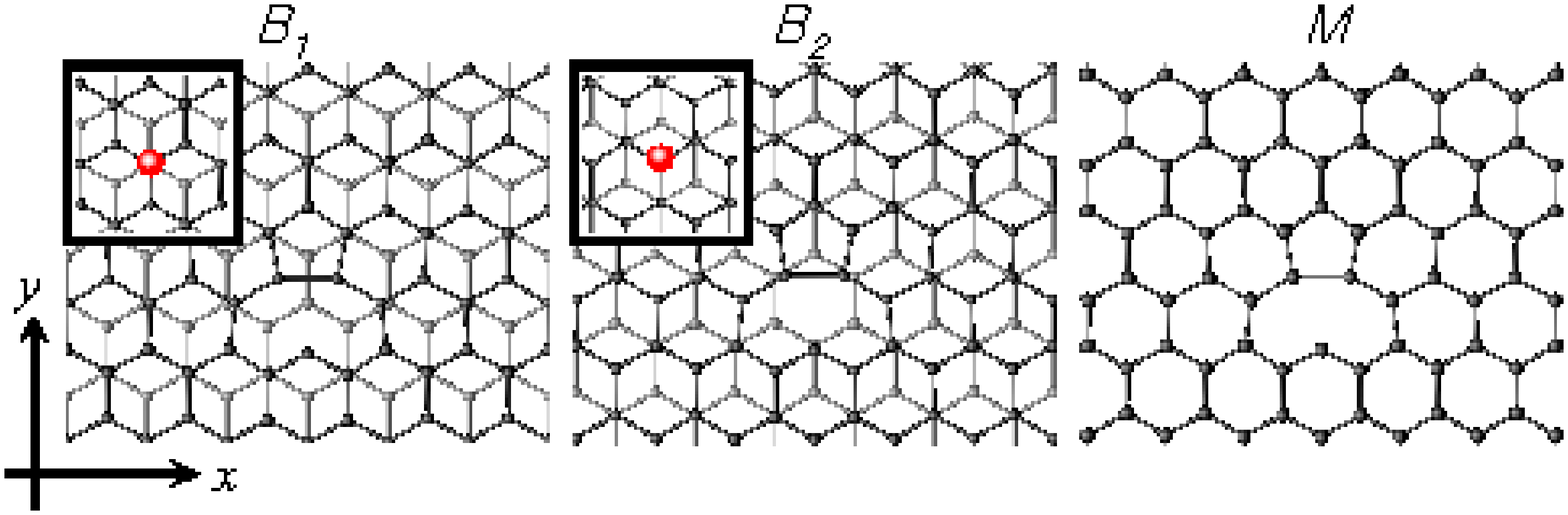}
\end{figure}
\begin{center}
\LARGE{Figure 1}

\LARGE{S. Choi et al.}
\end{center}

\newpage
\begin{figure}[t]
  \centering
  \includegraphics[width=11.0cm]{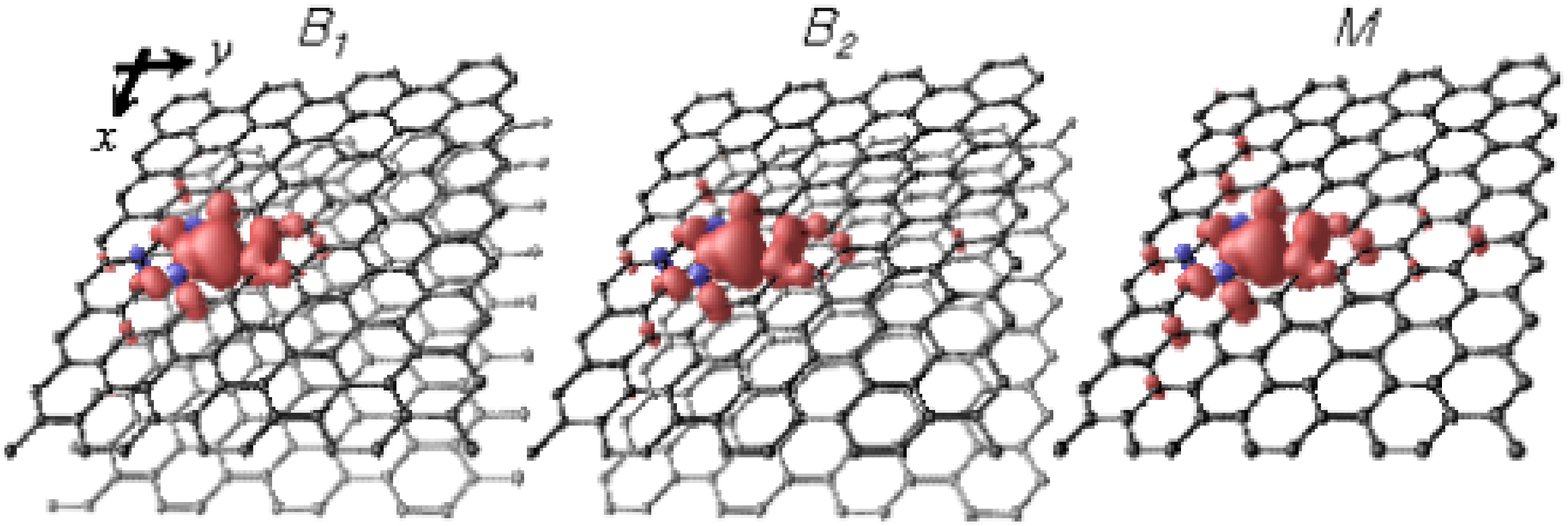}
\end{figure}
\begin{center}
\LARGE{Figure 2}

\LARGE{S. Choi et al.}
\end{center}

\newpage
\begin{figure}[t]
  \centering
  \includegraphics[width=11.0cm]{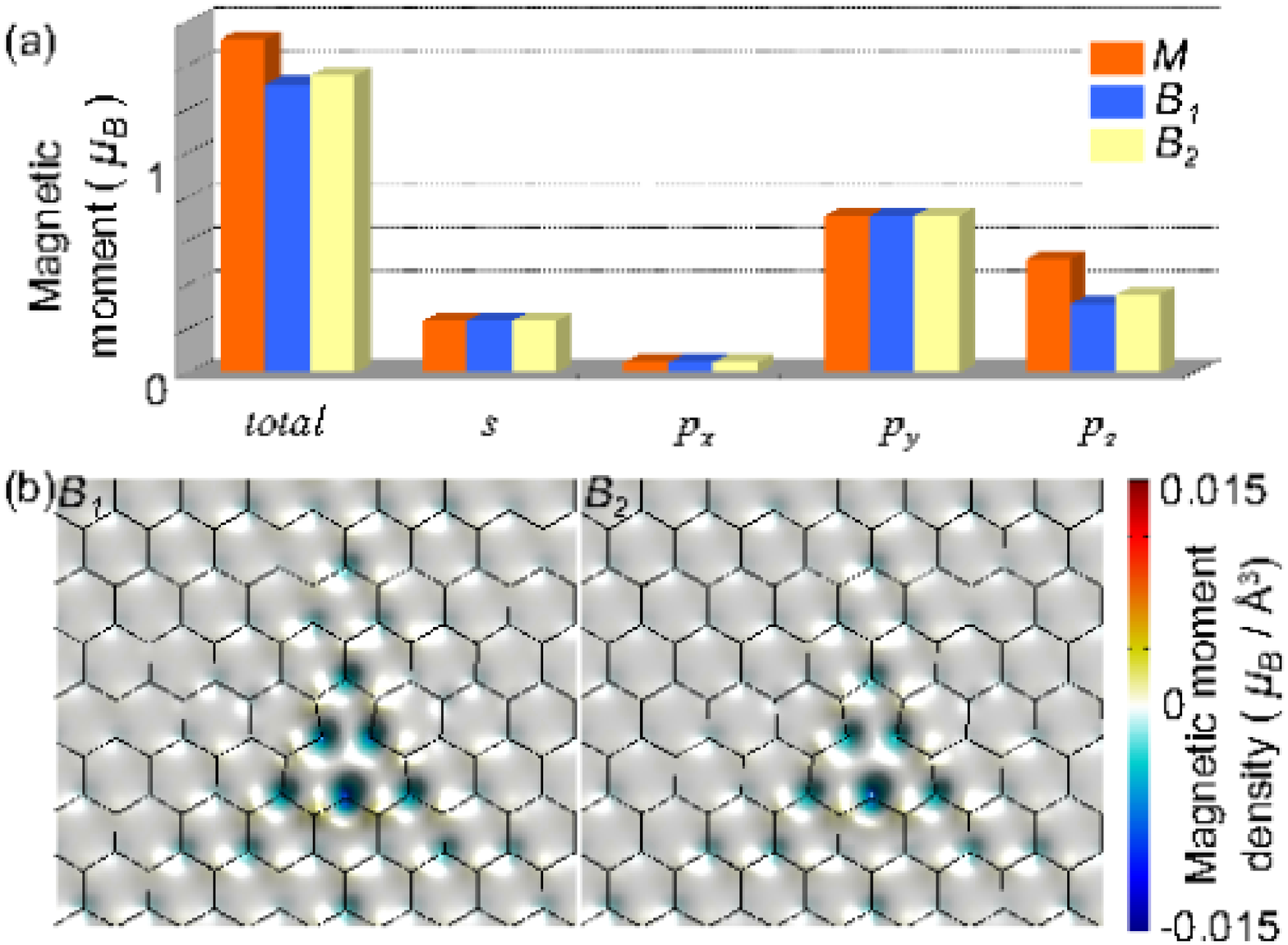}
\end{figure}
\begin{center}
\LARGE{Figure 3}

\LARGE{S. Choi et al.}
\end{center}

\newpage
\begin{figure}[t]
  \centering
  \includegraphics[width=11.0cm]{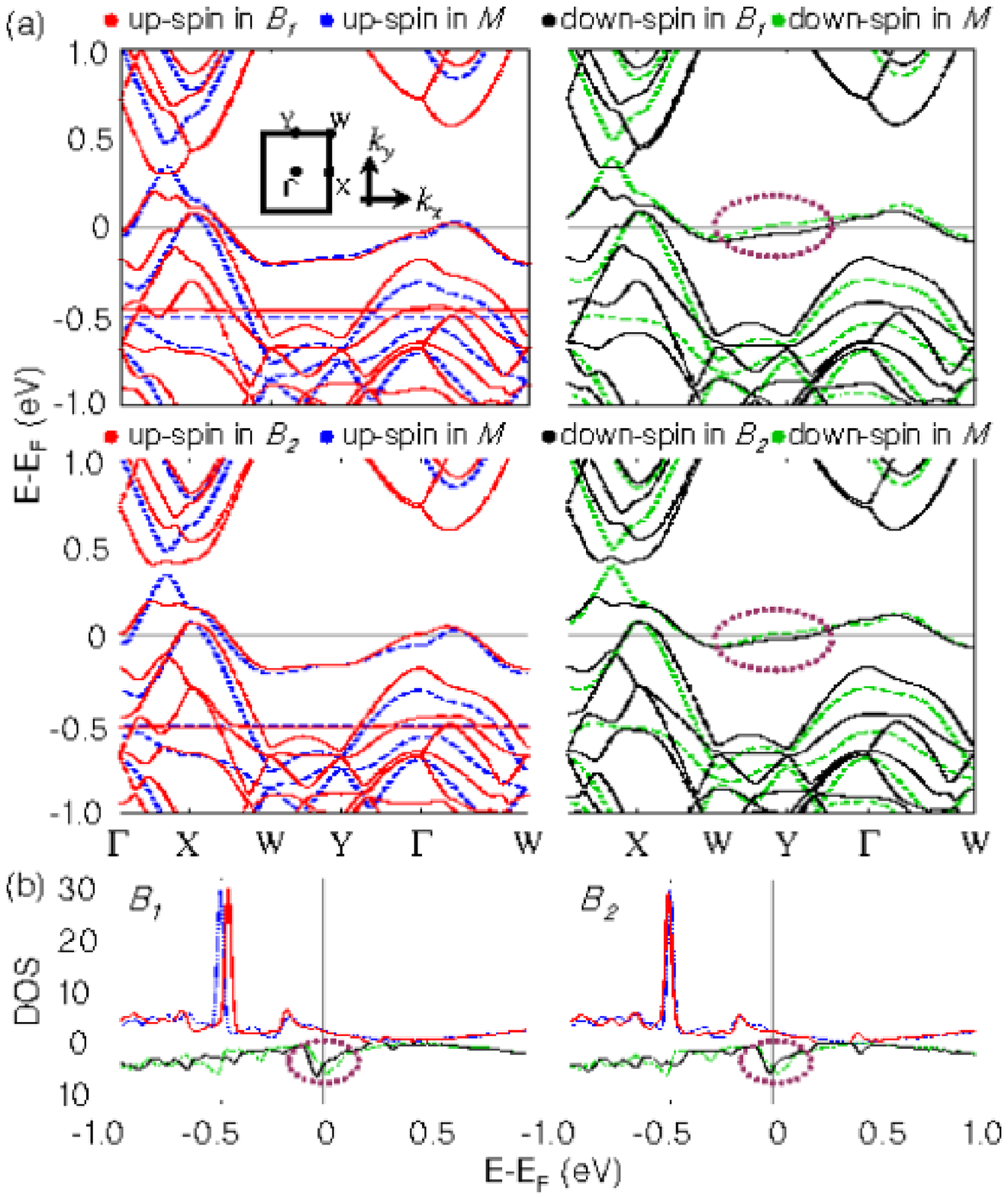}
\end{figure}
\begin{center}
\LARGE{Figure 4}

\LARGE{S. Choi et al.}
\end{center}

\end{document}